\documentclass[4p,12]{elsarticle}
\usepackage{verbatim}
\usepackage{lineno,hyperref}
\modulolinenumbers[5]
\usepackage{color}
\journal{}

\usepackage[]{subfigure}
\usepackage{placeins} 
\usepackage[english]{babel}  
\usepackage[utf8x]{inputenc}
\usepackage{graphicx}
\usepackage{booktabs}
\usepackage{tabularx}
\usepackage{caption}
\usepackage{amsmath}
\usepackage{amssymb}
\usepackage{multirow}
\usepackage{mathtools}

\let\today\relax
\makeatletter
\def\ps@pprintTitle{%
    \let\@oddhead\@empty
    \let\@evenhead\@empty
    \def\@oddfoot{\footnotesize\itshape
         {} \hfill}%
    \let\@evenfoot\@oddfoot
    }
\makeatother

\begin{document}
\small
\begin{frontmatter}

\title{Self-locking in Collapsed Carbon Nanotube Stacks via Molecular Dynamics}

\author[FBK]{Andrea Pedrielli}
\author[ECT,TIFPA]{Simone Taioli}
\author[DICAM,LONDON]{Nicola M. Pugno \corref{cor1}}

\address[FBK]{Fondazione Bruno Kessler, Trento, Italy}
\address[ECT]{European Centre for Theoretical Studies in Nuclear Physics and Related Areas (ECT*), Fondazione Bruno Kessler, Trento, Italy}
\address[TIFPA]{Trento Institute for Fundamental Physics and Applications (TIFPA-INFN), Trento, Italy}
\address[DICAM]{Laboratory for Bioinspired, Bionic, Nano, Meta Materials \& Mechanics, Department of Civil, Environmental and Mechanical Engineering, University of Trento, Trento, Italy}
\address[LONDON]{School of Engineering and Materials Science, Queen Mary University of London, UK}

\ead{pugno@unitn.it}
\cortext[cor1]{}

\begin{abstract}
Self-locking structures are often studied in macroscopic energy absorbers, but the concept of self-locking can also be effectively applied at the nanoscale. In particular, we can engineer self-locking mechanisms at the molecular level through careful shape selection or chemical functionalisation. The present work focuses on the use of collapsed carbon nanotubes (CNTs) as self-locking elements. We start by inserting a thin CNT into each of the two lobes of a collapsed larger CNT. We aim to create a system that utilises the unique properties of CNTs to achieve stable configurations and enhanced energy absorption capabilities at the nanoscale. We have used molecular dynamics simulations to investigate the mechanical properties of periodic systems realised with such units. This approach extends the application of self-locking mechanisms and opens up new possibilities for the development of advanced materials and devices.
\end{abstract}

\end{frontmatter}


\section{Introduction}

\begin{figure}[hbt!]
\centering
\includegraphics[width=1.0\textwidth]{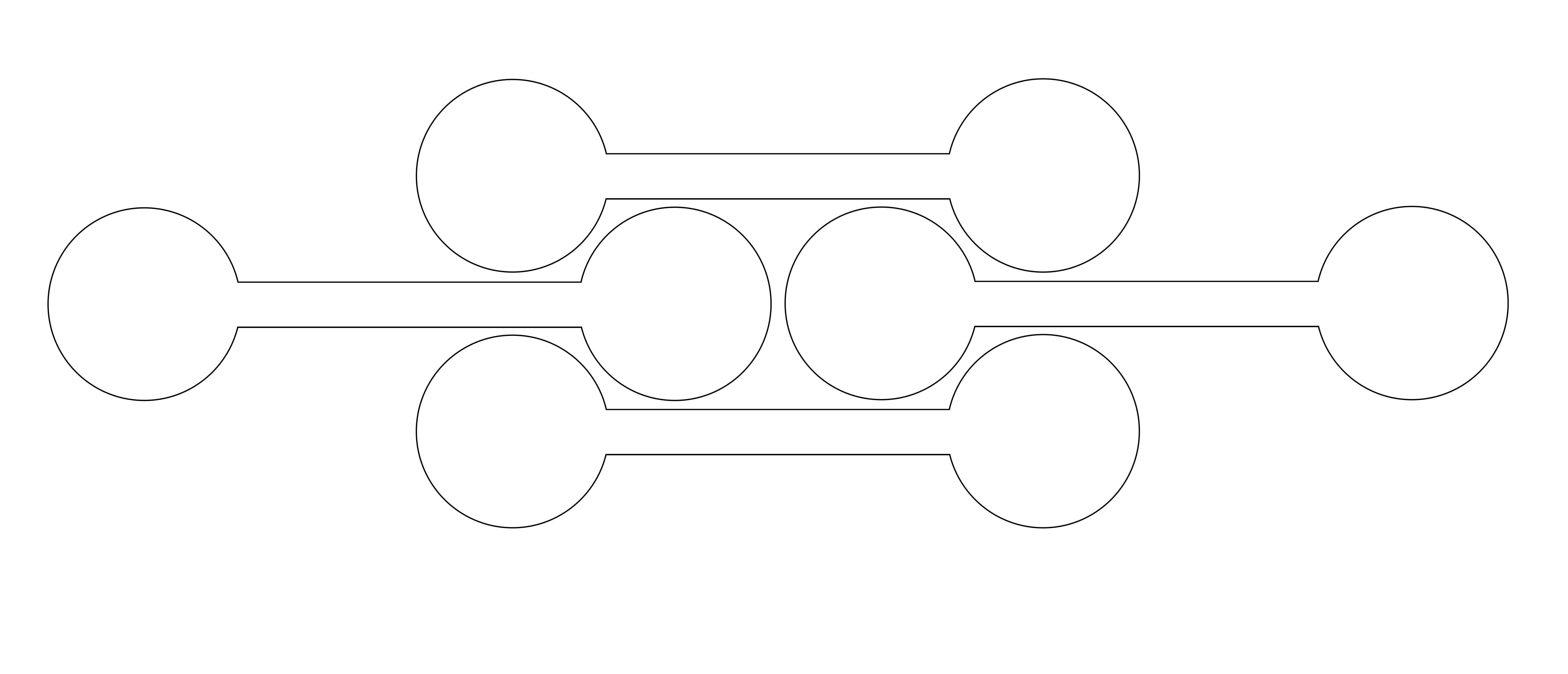}
\caption{Cross-sectional view of a self-locking arrangement of dumbbell-shaped tubes.}
\label{fig:Dumbbells}
\end{figure}

Thin-walled round tubes are commonly used as energy absorbers on a macroscopic scale \citep{Reid1978, Olabi2007, Baroutaji2017}. However, the structures based on stacks of round tubes are prone to splashing without adding internal connections or lateral restraints. Since the insertion of inter-tubes or lateral restraints may be undesirable or cumbersome to implement, various strategies have been proposed to avoid splashing. One strategy to avoid lateral splashing is based on elements that interlock to form self-locking structures. 
This concept has already been applied, for example, in the development of energy absorbers on a macroscopic scale \citep{Chen2016}.
In this context, various architectures have been developed to improve this energy absorption. One of the most promising designs, based on the dumbbell shape \citep{Chen2016, Qiao2017, Zhao2019}, is shown in Fig. \ref{fig:Dumbbells}.

However, in the development of nanostructures with novel functionalities, the use of self-locking strategies is still limited to certain niches \citep{Thomas2007, Nishio2015}. For example, self-assembled molecular structures are sometimes designed to be self-locking \citep{Du2023}. It would be obvious to extend this concept to other areas of nanoscience, such as the broad field of 2D materials.
At the nanoscale, the dumbbell shape can be easily found, for example, as a cross-section of a collapsed CNT \citep{Chopra1995}, which have been most studied for their remarkable mechanical \citep{Xiao2007, Pugno2010, Cheng2013, Meng2019, Jensen2020} and thermal \citep{AlGhalith2017, Qin2023} properties.

Following this route, self-locking materials based on collapsed CNTs could be developed.
While a stack of longitudinally aligned collapsed CNTs could be self-locking under low pressure, we could improve the self-locking behaviour by designing the unit block as a large collapsed CNT with thinner CNTs inserted (or grown) in both lobes, as shown in Fig. \ref{fig:BuildingBlock}. Indeed, thin CNTs exhibit extremely high mechanical properties (i.e. strength of $\approx 100$ GPa and Young's modulus of $\approx 1$ TPa for the $(5,5)$ CNTs \citep{Krishnan1998}).
The choice of the ratio between the radius of the outer nanotubes and that of the two supporting CNTs should enable the stable collapse of the central part of the large CNT. If the radius of the inner nanotubes is $R$ and $D=0.335$ nm is the interlayer spacing of graphite, the outer one should have a circumference of at least $C=2(2\pi(R+D))$ for geometric constraints. In the case of self-collapse, a longer C would be needed \citep{Pugno2010}. We note that these structures are far from being hypothetical, they were obtained at the experimental level \cite{Endo2004}. 

This self-locking building block could be further stabilised by inducing defects and chemical bonds between the inner CNTs and the larger collapsed CNTs. Similarly, bonding could be implemented within the entire stack of building blocks. We note that the methods for realising these connections at the nanoscale differ from those at the macroscale and could be profitably coupled with the concept of self-locking. There are various approaches in this respect, such as electron \citep{Kis2004} or ion beam irradiation \citep{Federizzi2006, OBRIEN2013}. A more recent approach is based on a combination of pulsed electric current, temperature and pressure \citep{Jensen2020}. In addition, the interposition of linkers between the CNTs \citep{PARK2017, Cornwell2012} could extend the applications of self-locking to chemically stabilised structures.  

In the present work, we implement the idea of self-locking at the nanoscale with CNTs of different sizes and determine the mechanical properties of self-locking structures based on collapsed CNTs with lobes supported by thin CNTs. We have used molecular dynamics (MD) simulations to track the atomic trajectories under compression and calculate the stress-strain curves and other mechanical properties to determine the dimensions for an optimal design of the self-locking architectures. MD simulations with appropriate force fields capable of accurately tracking the high strain regimes have indeed provided useful insights into the mechanical and thermal properties of CNTs and graphene-based structures \citep{Cornwell2012, OBRIEN2013, Pedrielli2017, Pedrielli2018}. However, unlike the standard evaluation of the mechanical properties of CNT bundles \cite{McCarthy2013, Mirzaeifar2015, Zhang2020, Pedrielli2023}, where most tests are performed with loads in the longitudinal direction, that is along the bundle axis, in this work we have focused on the mechanical properties under loading in the transverse direction, i.e. in the direction in which the nanotubes are collapsed.
We find that the use of different building blocks, determined by both the absolute value and the ratio between the outer and inner CNT radius, leads to a significantly different behaviour of the stress-strain curves and plays an important role in the self-locking mechanism. In particular, some combinations of outer and inner tubes actually show a complex collective behaviour under compression, which calls for further research on this topic.

\section{Materials and Methods}

\begin{figure}[htbp!]
\centering
\includegraphics[width=1.0\textwidth]{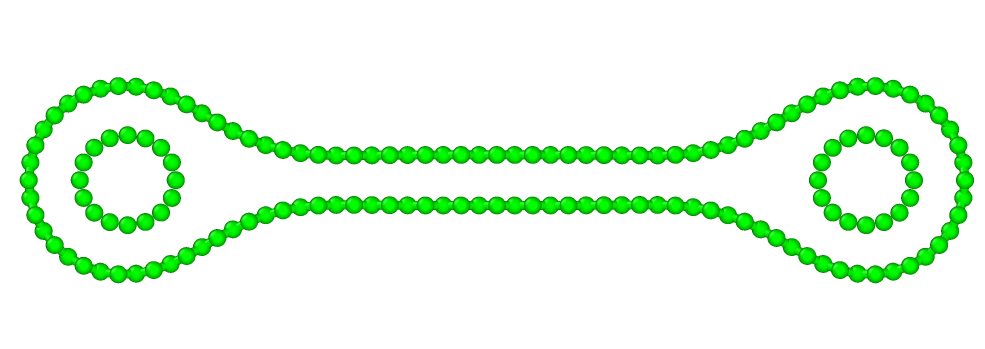}
\caption{Cross-sectional view of a collapsed $(60,0)$ CNT with $(8,0)$ supporting CNTs.}
\label{fig:BuildingBlock}
\end{figure}

Our building blocks were designed by assembling collapsed CNTs with thin CNTs inside the lobes as support. An illustration of such a building block can be found in Fig. \ref{fig:BuildingBlock}.

\begin{figure}[hbt!]
\centering
\includegraphics[width=0.15\textwidth]{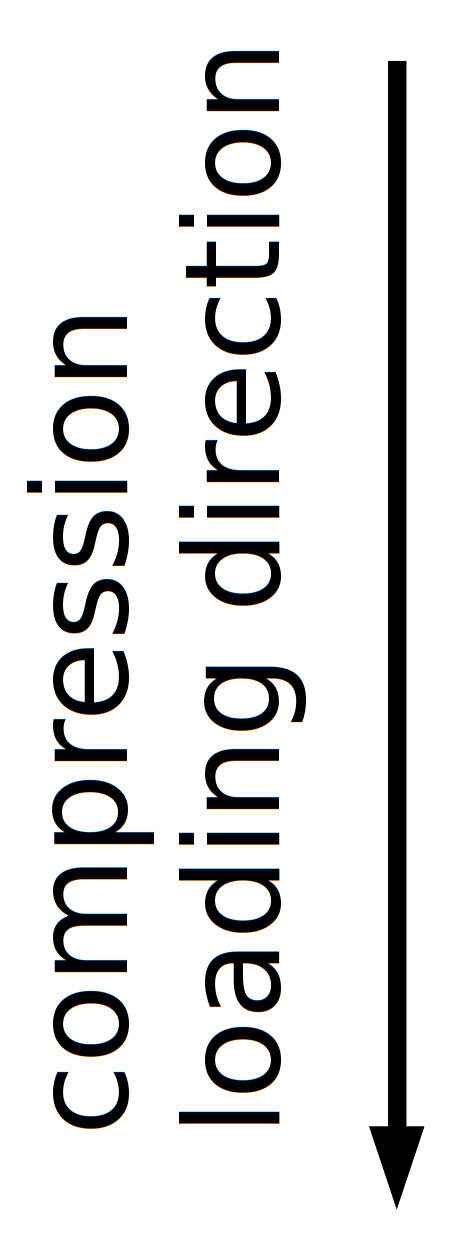}
\includegraphics[width=0.84\textwidth]{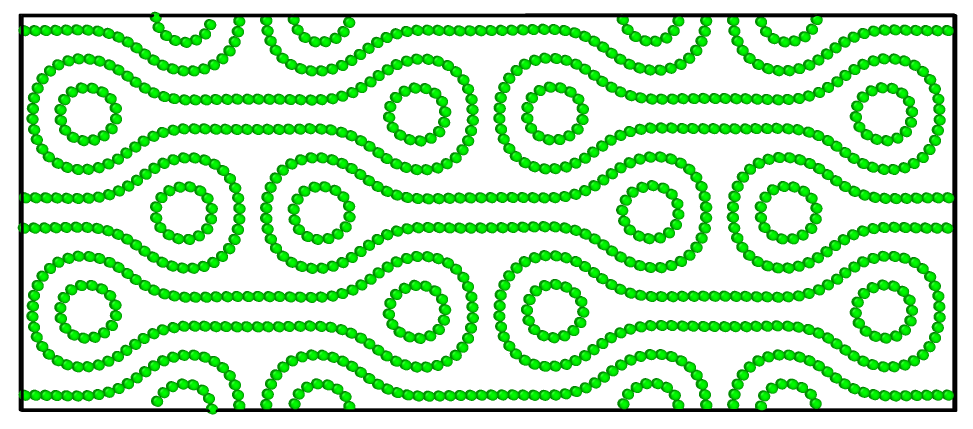}
\caption{Cross-sectional view of a periodic stack of a collapsed $(50,0)$ CNT with $(8,0)$ supporting CNTs, relaxed using molecular dynamics.}
\label{fig:RelaxedStack}
\end{figure}

First, we investigated the combination of collapsed $(50,0)$, $(60,0)$, $(70,0)$, $(80,0)$ zig-zag CNTs with $(8,0)$, $(10,0)$, $(12,0)$, $(14,0)$ CNTs inside the lobes. In the case of the largest collapsed CNT, we used the $(16,0)$ and the $(18,0)$ instead of the $(8,0)$ as supporting CNTs. The choice to focus on zig-zag CNTs is not particularly limiting, but this study could be extended to different combinations of CNT chiralities (despite the fact that zig-zag CNTs are simply the most stable at long length \citep{HEDMAN2017443}, chirality does not matter in this context).
The building blocks were arranged as in Fig. \ref{fig:Dumbbells}, which on the nanoscale leads to the arrangement shown in Fig. \ref{fig:RelaxedStack} for the case of a ($(50,0)$-$(8,0)$) stack. The simulation cell consists of a total of $8$ blocks in the plane, and periodic boundary conditions apply in the three orthogonal directions with 3 unitary cells that are periodically repeated in the out-of-plane direction.

The mechanical properties were calculated using the MD method implemented in the LAMMPS code \citep{Plimpton1995}. The calculations were performed with the interatomic potential AIREBO \citep{Stuart2000}.

The AIREBO potential energy consists of three terms
\begin{linenomath}
\begin{equation}
E  = \frac{1}{2} \sum_i \sum_{j \neq i}
\left[ E^{\text{REBO}}_{ij} + E^{\text{LJ}}_{ij} +
 \sum_{k \neq i,j} \sum_{l \neq i,j,k} E^{\text{TORSION}}_{kijl} \right],
\end{equation}
\end{linenomath}

where $i$, $j$, $k$, $l$ denote the atoms. The first term is the Reactive Empirical Bond Order (REBO) potential \cite{Brenner2002}, the second is the Lennard--Jones term for the non-bonded interaction, and the third is an explicit 4-body potential for the description of the dihedral angles.

The temperature was set to $T = 1$K and the pressure control in the directions transverse to the deformation was realised with the Nosé-Hoover barostat-thermostat damping coefficient, which was assumed to be $1$ in our simulations. The lateral pressure was set to a target value of $0$ Pa to simulate free boundaries. The velocity Verlet integrator with a time step of $1$ fs enables the correct integration of Newton’s equations of motion (i.e. the conservation of the total energy). Prior to the mechanical tests, all samples were fully equilibrated at $T = 1$K and a pressure of zero.

The simulations were performed by forcing a uniform contraction of the simulation cell in the vertical direction and remapping the atomic positions (see the following Figs. \ref{fig:CompressionPhases}, \ref{fig:SymmetricToAsymmetric}, \ref{fig:SeparatedLobes}). We evaluated the mechanical properties by calculating the total stress of the system and in particular the component of the total stress along the direction of compression. The strain parallel to the direction of deformation is defined by:
\begin{linenomath}
\begin{equation}
\varepsilon = \frac{L-L_0}{L} = \frac{\Delta L}{L},
\end{equation} 
\end{linenomath}
where $L_0$ and $L$ are the initial and actual length of the sample in the direction of loading. 

The uniaxial compressive load was applied until a total strain of $40$\% was reached. The applied strain rate was set to $0.1$\% per ps. Stress and strain were stored every $1000$ time steps.

To determine the stress, the components of the compressive stress tensor in response to the external deformation are calculated as \citep{Thompson2009}
\begin{equation}\label{pressure}
P_{ij} = \frac{\sum_k^N{m_k v_{k_i} v_{k_j}}}{V}+ \frac{\sum_k^N{r_{k_i} f_{k_j}}}{V},
\end{equation}
where $i$ and $j$ denote the coordinates $x$, $y$, $z$; $k$ runs over the atoms; $r_{k_i}$, $m_k$ and $v_k$ are the position, mass and velocity of the $k$th atom; $f_{k_j}$ is the $j$th component of the total force on the $k$th atom due to the other atoms; and finally $V$ is the volume of the simulation box.

The performance of our stacks as energy absorbers was assessed by calculating the specific energy absorption (SAE):
\begin{equation}\label{sae1}
\mathrm{SAE}=\frac{E_t}{M},
\end{equation}
where $M$ is the mass of the sample and $E_t$ is the total absorbed energy defined as follows:

\begin{equation}
E_t= V \int_{0}^{\varepsilon}\sigma(\varepsilon')\text{d}\varepsilon',
\end{equation} 

where $\sigma(\varepsilon')$ is the stress at $\varepsilon'$ strain, and $V$ is the sample volume. The upper integration limit was set at $40$\% strain or the strain at which the stacks fail abruptly.

Snapshots of the atomic configurations were taken with the OVITO package \citep{Stukowski2009}.
The von Mises stress was calculated as follows
\begin{equation}
\sigma_\text{VM} = \sqrt{\frac{1}{2}\left[\left(\sigma_{11} - \sigma_{22}\right)^2 + 
\left(\sigma_{22} - \sigma_{33}\right)^2 +
\left(\sigma_{33} - \sigma_{11}\right)^2 \right] +
3 \left(\tau^2_{12} + \tau^2_{23} + \tau^2_{31}\right) },
\end{equation}
where $\sigma$ and $\tau$ are the normal and the shear stresses, respectively.
The von Mises stress was visualised using the same colour code for the atomic configurations at different stages of compression, i.e. blue (zero stress) and red (maximum stress).

\section{Results}

\begin{figure}[hbt!]
\centering
\includegraphics[width=1.0\textwidth]{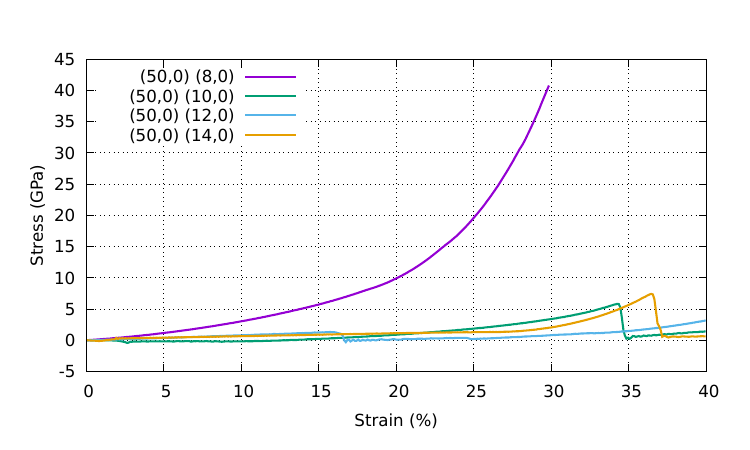}
\caption{Stress-strain curves of a stack of collapsed $(50,0)$ CNTs and various supporting CNTs ($(8,0)$, $(10,0)$, $(12,0)$, $(14,0)$). Periodic boundary conditions were applied to the simulation cell in all directions.}
\label{fig:Collapsed50}
\end{figure}

In Fig. \ref{fig:Collapsed50} we report the stress-strain curves of of a periodic stack of collapsed $(50,0)$ CNTs and various supporting CNTs ($(8,0)$, $(10,0)$, $(12,0)$, $(14,0)$). 
\\
According to our simulations, the mechanical failure of these systems can be explained in two ways. The first type of mechanical failure occurs when the structure is strongly self-locking and the failure is caused by a single component. In this case, we have plotted the stress-strain curve only up to the maximum stress before mechanical failure. The second type of failure is due to sliding between different building blocks in the direction perpendicular to the direction in which the pressure is applied. In this case, we plot the entire stress-strain curve as the energy release is limited and the simulations are still stable after unlocking.

\begin{figure}[hbt!]
\centering
\includegraphics[height=0.4\textwidth]{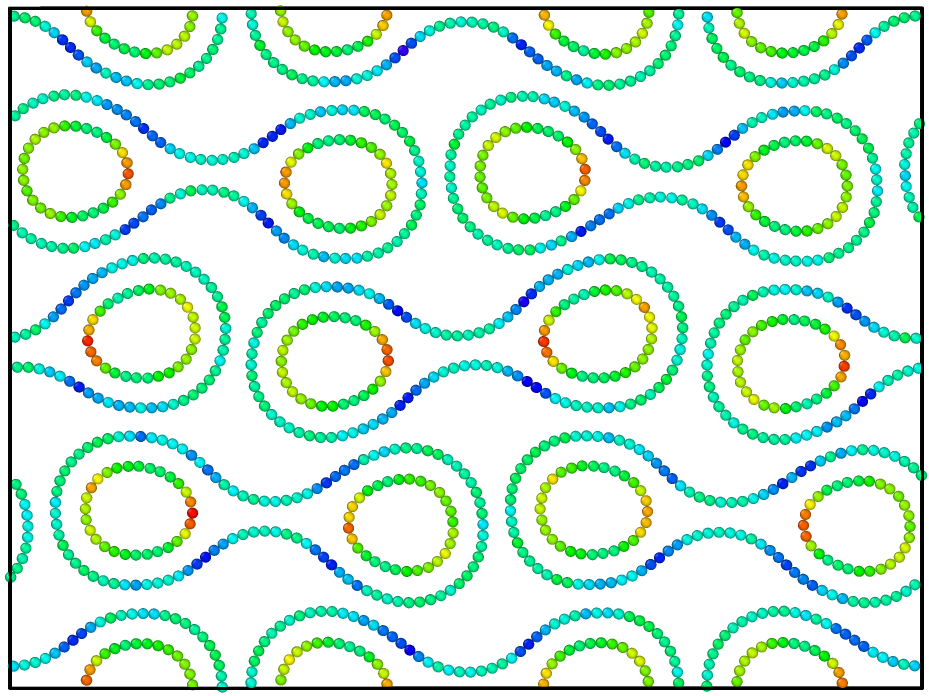}\\$~$\\
\includegraphics[height=0.36\textwidth]{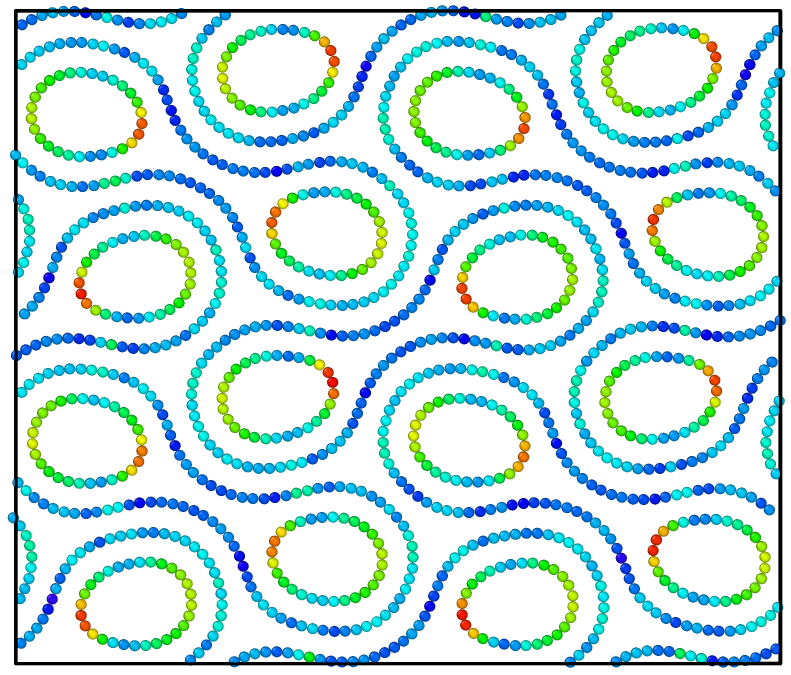}\\$~$\\
\includegraphics[height=0.28\textwidth]{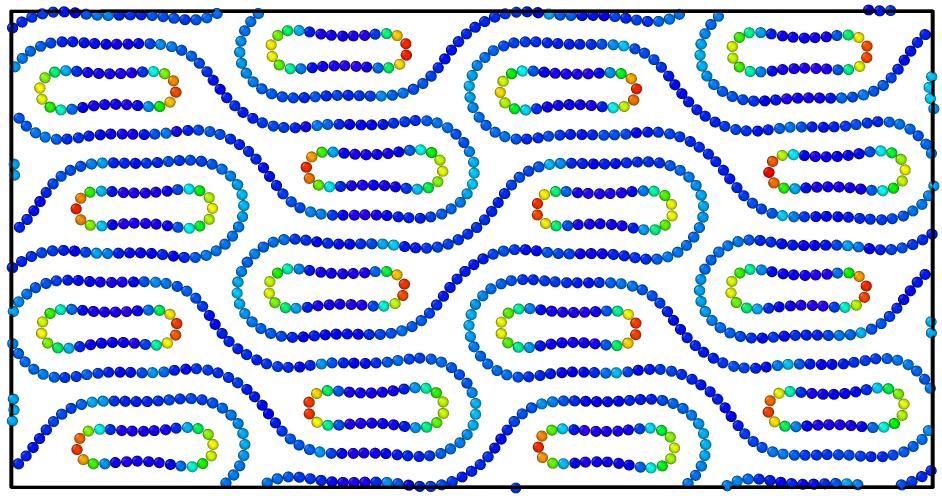}\\$~$\\
\includegraphics[height=0.24\textwidth]{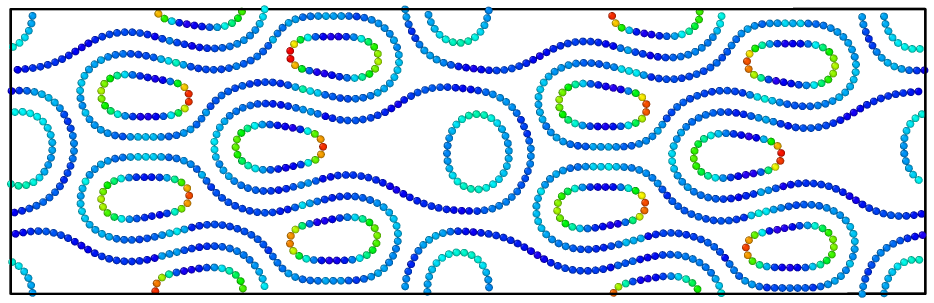}
\caption{Snapshots of the atomic configurations for the $(50,0)-(14,0)$ stack at different strain. The upper panel shows the almost symmetric relaxed configuration. The second panel from the top shows the asymmetric configuration at $10$\% strain. The configuration at $30$\% strain is shown in the third panel. Finally, the configuration at $40$\% strain is shown in the bottom panel.}
\label{fig:CompressionPhases}
\end{figure}

As you can see, there is a significant difference between the behaviour of the stack when using the inner CNTs $(8,0)$ with the smallest diameter and the other cases. In this case, the self-locking is very efficient and the system is stable under compression, allowing a progressive increase in stress during compression, that corresponds to a stiffening of the material. The failure of the system is caused by a mechanical failure of the single building block. In such case, as already mentioned, we have plotted the stress-strain curve at this specific maximum strain before mechanical failure.
In the second case $(50,0)$-$(10,0)$, the stress initially decreases and becomes negative due to the equilibrium between the elastic energy and the interlayer energy. The stress then tends to increase. At $34$\% strain, the stack collapses.
We note that the scenario for the $(50,0)$-$(12,0)$ stack is somewhat complementary to that of the sample built with $(50,0)$-$(10,0)$ CNTs. Indeed, we have reached a stable minimum during the initial relaxation, but the structure loses self-locking at relatively low strain ($17$\%).
For the structure with the larger inner CNTs $(50,0)$-$(14,0)$, the structure initially rearranges, followed by a slow increase in stress, and the stack loses self-locking at $37$\% strain.
Four snapshots of the atomic configurations for the $(50,0)$-$(14,0)$ stack at different strains are shown in Fig. \ref{fig:CompressionPhases}. The upper panel shows the almost symmetrical relaxed configuration. Even if this configuration is locally stable, it is unstable under compression. The second panel below shows the asymmetric configuration at $20$\% strain, where the structure is more compact. In the third panel from the top, you can see the still asymmetric configuration at a strain of 30\%, and in the bottom panel the configuration at a strain of 40\% where some of the lobes are no longer unlocked.

\begin{figure}[hbt!]
\centering
\includegraphics[width=1.0\textwidth]{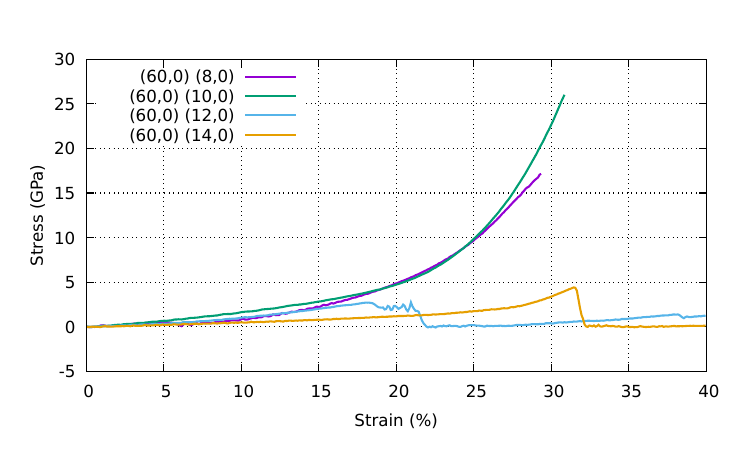}
\caption{Stress-strain curves of a stack of collapsed $(60,0)$ CNTs with different support CNTs ($(8,0)$, $(10,0)$, $(12,0)$, $(14,0)$). Periodic boundary conditions are applied to the simulation cell in all directions.}
\label{fig:Collapsed60}
\end{figure}

\begin{figure}[hbt!]
\centering
\includegraphics[height=0.3\textwidth]{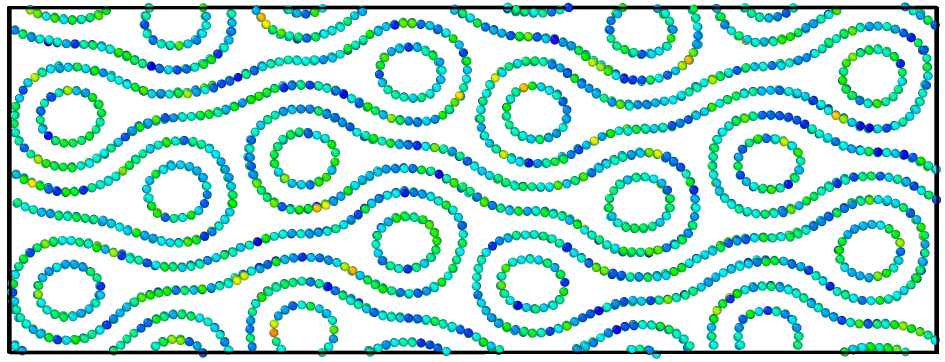}\\$~$\\
\includegraphics[height=0.233\textwidth]{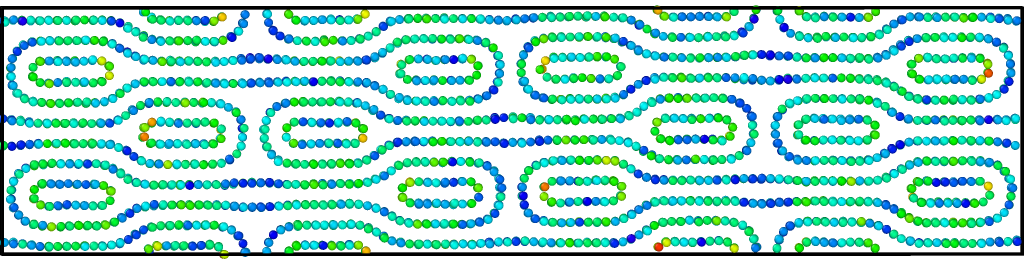}
\caption{Snapshots of the atomic configurations for the $(50,0)$-$(10,0)$ stack at different strain. The top panel shows the scaled snapshot for the asymmetric configuration at $10$\% strain and the bottom panel shows the symmetric configuration at $30$\% strain.}
\label{fig:SymmetricToAsymmetric}
\end{figure}

In Fig. \ref{fig:Collapsed60} we show the stress-strain curves of a periodic stack of collapsed $(60,0)$ CNTs and various supporting CNTs ($(8,0)$, $(10,0)$, $(12,0)$, $(14,0)$).
For the $(60,0)$-$(8,0)$ sample, we have a small rearrangement that brings the self-locking elements into an asymmetric configuration, which however seems to be quite stable. Similarly, the stack $(60,0)$-$(10,0)$ is stable and goes from an asymmetric self-locking configuration to a symmetric one. We have shown in Fig. \ref{fig:SymmetricToAsymmetric} this transition.
The structure consisting of $(60,0)$-$(12,0)$-CNTs is again asymmetric and loses self-locking at $21$\% strain.
For the $(60,0)$-$(14,0)$, we again obtain an asymmetric structure in the direction of compression. The stress increases evenly up to $32$\% strain when the self-locking is lost.


\begin{figure}[hbt!]
\centering
\includegraphics[width=1.0\textwidth]{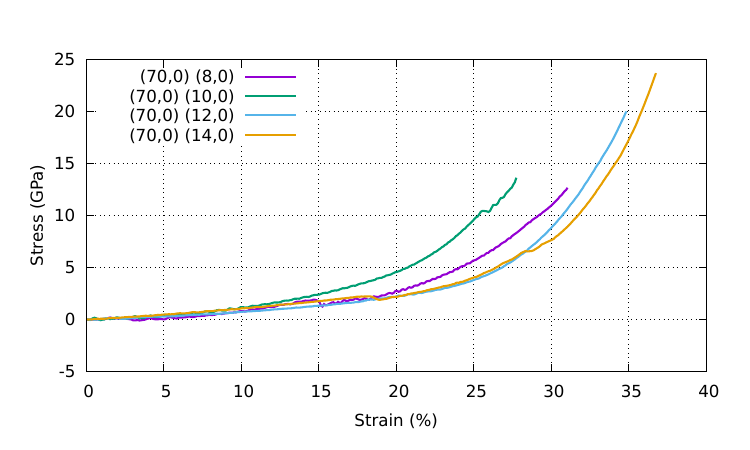}
\caption{Stress-strain curves of a stack of collapsed $(70,0)$ CNTs and various supporting CNTs ($(8,0)$, $(10,0)$, $(12,0)$, $(14,0)$). Periodic boundary conditions are applied in all directions.}
\label{fig:Collapsed70}
\end{figure}

In Fig. \ref{fig:Collapsed70} we show the stress-strain curves of a periodic stack of collapsed $(70,0)$ CNTs and various supporting CNTs ($(8,0)$, $(10,0)$, $(12,0)$, $(14,0)$).
In the sample $(70,0)$-$(8,0)$, as in the previous case, we have a small rearrangement that brings the self-locking elements into an asymmetric configuration. This configuration is stable up to $15$\% strain when another small rearrangement occurs. The new configuration is still self-locking until the stack fails at $31$\% strain.
The $(70,0)$-$(10,0)$ specimen is self-locking with an asymmetrical arrangement and can gradually bear the load up to a strain of $28$\%.
The $(70,0)$-$(12,0)$ and $(70,0)$-$(14,0)$ stacks are self-locked with a small asymmetry in arrangement, their deformation under compression, and the maximum stress is higher than those of the other two samples.


\begin{figure}[hbt!]
\centering
\includegraphics[width=1.0\textwidth]{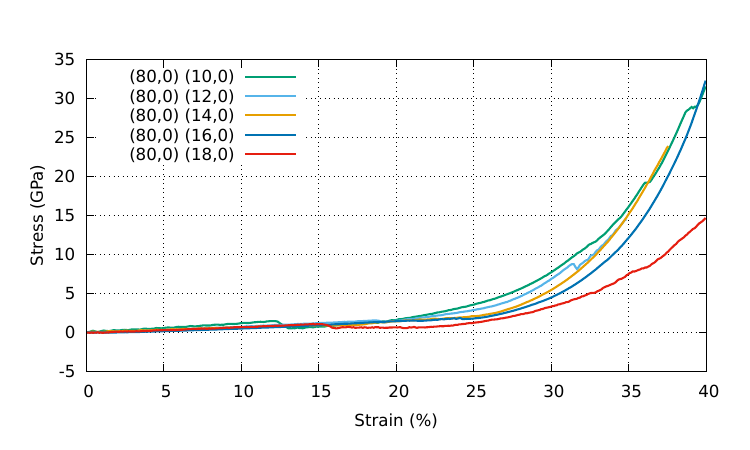}
\caption{Stress-strain curves of a stack of collapsed $(80,0)$ CNTs and various supporting CNTs ($(10,0)$, $(12,0)$, $(14,0)$, $(16,0)$, $(18,0)$). Periodic boundary conditions are applied to the simulation cell in all directions.}
\label{fig:Collapsed80}
\end{figure}

\begin{figure}[htbp!]
\centering
\includegraphics[height=0.2\textwidth]{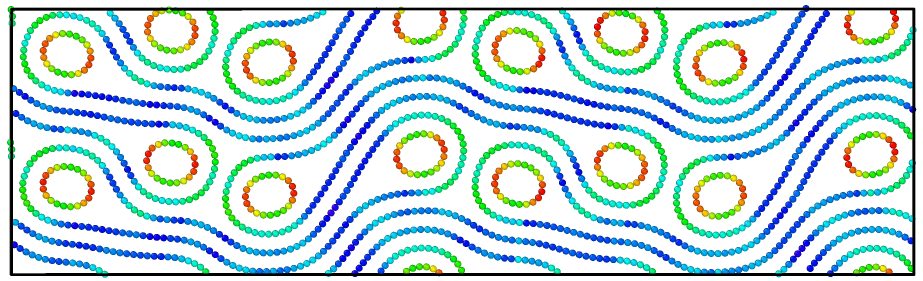}\\$~$\\
\includegraphics[height=0.1555\textwidth]{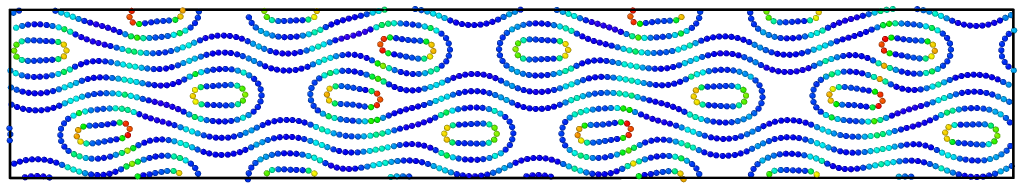}
\caption{Snapshots of the atomic configuration for the $(80,0)$-$(10,0)$ stack at different strains. The top panel shows the snapshot for the asymmetric configuration at $10$\% strain and the bottom panel shows the still asymmetric configuration at $30$\% strain.}
\label{fig:SeparatedLobes}
\end{figure}

In Fig. \ref{fig:Collapsed80} we have shown the stress-strain curves of a periodic stack of collapsed $(80,0)$ CNTs and different supporting CNTs ($(10,0)$, $(12,0)$, $(14,0)$, $(16,0)$, $(18,0)$).
For the sample $(80,0)$-$(10,0)$, there is a linear increase in stress that brings the system into a different asymmetric configuration at $12.5$\% strain. Beyond this point, the system is self-locking and stable. Fig. \ref{fig:SeparatedLobes} shows two snapshots of the atomic configurations for the $(80,0)$-$(10,0)$ stack at different strains. The top panel shows the snapshot for the asymmetric configuration at $10$\% strain and the bottom panel reports the snapshot for $30$\% strain. It is worth noting that in the case of $30$\% strain, a horizontal separation occurs between the lobes of the building blocks on the same plane. For the samples $(80,0)$-$(12,0)$ and $(80,0)$-$(14,0)$ we show interrupted stress-strain curves since the systems exhibit mechanical failure. In contrast, the strain is not sufficient to cause the stacks $(80,0)$-$(16,0)$ and $(80,0)$-$(18,0)$ to fail, although the behaviour is similar in both cases. 

In Tab. \ref{tab:Stress} we summarise the fracture and unlocking stresses. We summarise the fracture and unlocking strains in Tab. \ref{tab:FractureStrain}.

\begin{table*}[htbp]
  \centering
    \begin{tabular}{cccccccc}
    \hline
    \multicolumn{2}{c}{\multirow{2}[4]{*}{}} & \multicolumn{5}{c}{inner} \\
\cline{3-8}    \multicolumn{2}{c}{} & (8,0)   &   (10,0)    &    (12,0)     &  (14,0) &    (16,0)     &  (18,0) \\
    \hline
    \multirow{2}[10]{*}{\rotatebox{90}{outer}} & (50,0) & 40.2 &  5.83  & 0.73   & 7.43  & -  &  -     \\
\cline{3-8}   &     (60,0) & 17.1 &  25.6  &  2.74  & 4.43 &  - &  -  \\ 
\cline{3-8}    &     (70,0) & 12.4 & 13.2 & 19.7  & 23.3 &  - & -   \\ 
\cline{3-8}      &  (80,0) & - &  $>$ 31.5  & 14.4  &  22.5 & $>$ 32.3  & $>$ 14.7  \\ 	
    \hline
    \end{tabular}
  \caption{Fracture or unlock stress for the various stacks.}
  \label{tab:Stress}
\end{table*}

\begin{table*}[htbp]
  \centering
    \begin{tabular}{cccccccc}
    \hline
    \multicolumn{2}{c}{\multirow{2}[4]{*}{}} & \multicolumn{5}{c}{inner} \\
\cline{3-8}    \multicolumn{2}{c}{} & (8,0)   &   (10,0)    &    (12,0)     &  (14,0) &    (16,0)     &  (18,0) \\
    \hline
    \multirow{2}[10]{*}{\rotatebox{90}{outer}} & (50,0) & 29.8 &  34.3  & 16.6   & 36.5  & -  &  -     \\
\cline{3-8}   &     (60,0) & 29.3 &  30.8  &  21.0  & 31.5 &  - &  -  \\ 
\cline{3-8}    &     (70,0) & 31.0 & 27.7 & 34.8  & 36.7 &  - & -   \\ 
\cline{3-8}      &  (80,0) & - &  $>$ 40.0  & 34.8  &  37.2 & $>$ 40.0  & $>$ 40.0  \\ 	
    \hline
    \end{tabular}
  \caption{Fracture or unlock percentage strain for the various stacks.}
  \label{tab:FractureStrain}
\end{table*}

In Tab. \ref{tab:SAE} we give the calculated SAE (see Eq. \ref{sae1}) for the stacks up to a strain of $40$\%. For those where there is an abrupt failure of the individual building block and no unlocking, the SAE is given up to the value of the failure strain.

\begin{table*}[htbp]
  \centering
    \begin{tabular}{cccccccc}
    \hline
    \multicolumn{2}{c}{\multirow{2}[4]{*}{}} & \multicolumn{5}{c}{inner} \\
\cline{3-8}    \multicolumn{2}{c}{} & (8,0)   &   (10,0)    &    (12,0)     &  (14,0) &    (16,0)     &  (18,0) \\
    \hline
    \multirow{2}[10]{*}{\rotatebox{90}{outer}} & (50,0) & 2.74 &  0.27  & 0.20   & 0.38  & -  &  -     \\
\cline{3-8}   &     (60,0) & 0.74 &  1.01  &  0.24  & 0.25 &  - &  -  \\ 
\cline{3-8}    &     (70,0) & 0.63 & 0.70 & 1.16  & 1.28 &  - & -   \\ 
\cline{3-8}      &  (80,0) & - &  $>$ 1.81  & 0.73  &  0.96 &  $>$ 1.19  &   $>$ 0.79  \\ 	
    \hline
    \end{tabular}
  \caption{The values of the SAE in MJ Kg$^{-1}$ are given for the various stacks.}
  \label{tab:SAE}
\end{table*}



\section{Discussion}

As the results show, the behaviour of the collapsed CNT stacks under compression strongly depends on the relative size of the inner and outer CNTs. In addition, the absolute dimensions of the CNTs play a role, so that, for example, space is created between the lobes during compression, as in the $(80,0)$-$(10,0)$ stack.
Failure of the system can occur due to the abrupt release of potential energy stored in the system or due to a softer unlocking of the building blocks due to a slip in the direction transverse to the direction of compression.
We have found that there are different regimes under compression. If there is no spatial gap between the building blocks and the structure is symmetrical, the compression is uniform in the direction of self-locking. If the initial configuration is symmetric but shows some spatial gaps between the collapsed CNTs, the structure leans towards an asymmetric state. In addition, some stacks are asymmetric from the beginning but tend to be symmetric under high compressive loading. Finally, in the case of large outer collapsed CNTs and thin inner CNTs, we found that some space is created between the lobes of different collapsed CNTs on the same plane. As for the self-locking properties and symmetry, we summarise the data in Tab \ref{tab:Stability} and Tab. \ref{tab:Symmetry}.

\begin{table*}[htbp]
  \centering
    \begin{tabular}{cccccccc}
    \hline
    \multicolumn{2}{c}{\multirow{2}[4]{*}{}} & \multicolumn{5}{c}{inner} \\
\cline{3-8}    \multicolumn{2}{c}{} & (8,0)   &   (10,0)    &    (12,0)     &  (14,0) &    (16,0)     &  (18,0) \\
    \hline
    \multirow{2}[10]{*}{\rotatebox{90}{outer}} & (50,0) & L & 	U & U  & U & -  &  -     \\
\cline{3-8}   &     (60,0) & L &  L  & U &  U &  - &  -  \\ 
\cline{3-8}    &     (70,0) & L & 	L  & L  & L &  - & -   \\ 
\cline{3-8}      &  (80,0) & - &  L  &  L  &  L & L  & L   \\ 	
    \hline
    \end{tabular}
  \caption{The property of being self-locking (L) until abrupt failure (at $40$\% strain) or unstable and unlocking (U) is summarised for the different stacks.}
  \label{tab:Stability}
\end{table*}

\begin{table*}[htbp]
  \centering
    \begin{tabular}{cccccccc}
    \hline
    \multicolumn{2}{c}{\multirow{2}[4]{*}{}} & \multicolumn{5}{c}{inner} \\
\cline{3-8}    \multicolumn{2}{c}{} & (8,0)   &   (10,0)    &    (12,0)     &  (14,0) &    (16,0)     &  (18,0) \\
    \hline
    \multirow{2}[10]{*}{\rotatebox{90}{outer}} & (50,0) & S & 	A & A  & A & -  &  -     \\
\cline{3-8}   &     (60,0) & A &  A $\xrightarrow{}$  S  & A & A  &  - &  -  \\ 
\cline{3-8}    &     (70,0) & A & 	A  & A  & A $\xrightarrow{}$ S &  - & -   \\ 
\cline{3-8}      &  (80,0) & - &  A  &  A  &  A & A $\xrightarrow{}$ S  & A   \\ 	
    \hline
    \end{tabular}
  \caption{The property of being symmetrical (S) or asymmetrical (A) in relation to the compression axis is summarised for the different stacks. The arrows indicate the cases in which the symmetry has changed due to the compression.}
  \label{tab:Symmetry}
\end{table*}

As for the shape of the stress-strain curves, the slope of the curves increases with increasing compression, this corresponds to a stiffening of the material under compression. We note that this is different from, for example, the compression of 3D structures such as carbon nanotruss networks \citep{Pedrielli2017} and carbon nanofoams \citep{Pedrielli2018}. In these cases, the stress-strain curve initially shows a steep increase, followed by a plateau during the collapse of the structure, and finally, a steep increase in stress when the structure is fully compressed.

We also note some differences from the macroscopic case of, for example, steel or PLA dumbbell tubes \citep{Chen2016, Zhao2019}. In the macroscopic case for the dumbbell-shaped tube, the angle between the round and the flat part (Fig. \ref{fig:Dumbbells}) tends to remain constant under compression, as the steel is shaped with this angle from the beginning and the tube opens in the center, effectively assuming a three-lobed shape. In the case of our systems, the bending stiffness of graphene is low and there is no similar effect. 
In addition, the stress-strain curves for the self-locking stacks are almost linear for the structures realised at the macroscopic level with dumbbell-shaped tubes \citep{Chen2016, Zhao2019}, while they are non-linear for our samples and show stiffening under compression. We note that the linearity of the stress-strain curve at the macroscopic level could be due to the design. In fact, similar results were obtained for steel and PLA structures \citep{Chen2016, Zhao2019}.

\FloatBarrier

\section{Conclusions}

In this work, we used molecular dynamics to investigate the mechanical properties of self-locking stacks based on collapsed CNTs. We have shown that the self-locking properties can be improved by inserting thinner CNTs into the lobes of the collapsed CNTs.

We have analysed in detail the atomic configurations of the structures corresponding to the different regimes shown in the stress-strain curves.
Our study shows that the building blocks used in the present work lead to a complex collective behaviour.
For certain combinations of the dimensions of the outer and inner CNTs, we obtained self-locking stacks. The ratio between the outer and inner CNT dimensions plays a role in self-locking, as does the absolute dimension of the same components.

In general, we found that self-locking can be achieved in most of the stacks analysed. However, only in a few cases, we found self-locking without rearrangement with respect to the relaxed initial configuration. In particular, the combinations of CNTs that provided us with the smoother stress-strain curves were $(50,0)$-$(8,0)$, $(60,0)$-$(10,0)$ and $(70,0)$-$(10,0)$. 
Applications that could utilise the self-locking mechanism of these nanostructures under pressure are conceivable in electronics \citep{Giusca2007}, thermodynamics \citep{AlGhalith2017} and mechanics \citep{Jensen2020}. For example, the electronic properties can be tuned by applying pressure, whereby a mechanically stable structure is maintained even without chemical contact between the collapsed tubes. In thermal applications, the ultra-high thermal conductivities of the individual CNTs are maintained during the collapse \citep{AlGhalith2017}, making these self-locking stacks suitable for fibers with high thermal conductivity and high mechanical stability. In addition, the thermal conductivity can be adjusted in the transverse direction by applying a load to the stack. The self-locking would be also a useful way to have nanotube cables resistant to transverse separation. We also devise that self-locking mechanisms could be widely applied in 2D layered materials, opening up new possibilities for developing advanced materials and devices.

Finally, future prospects and further investigations could consider e.g. increasing the size of the building blocks by using multilayer CNTs or connecting the nanotubes by e.g. single carbon atom linkers.
On the computational side, different force fields could be used to study specific regimes. For example, further insights could be gained by using the AIREBO-M \citep{OConnor2015} force field, which is valid for high-pressure regimes, or by replacing the long-range interaction with potentials to better treat the interactions between the layers, such as the dihedral-angle-corrected registry-dependent potential (DRIP) \citep{Wen2018}, which can also be coupled with potentials that can treat bonded interactions, such as REBO \citep{Stuart2000} or Tersoff \citep{Tersoff1988}.

\section*{Acknowledgements}
This action has received funding from the European Union under grant agreement No 10104665

\bibliography{Manuscript}

\end{document}